\documentclass[preprint,prl,showpacs]{revtex4}
\usepackage{natbib}
\usepackage{graphicx}
\usepackage{epsfig}
\usepackage{ae}
\usepackage{aecompl}

\begin{document}
\bibliographystyle{apsrev}

\title{Phase diagram of a ternary mixture of cholesterol and saturated
and unsaturated lipids calculated from a microscopic model}

\author{R. \ Elliott$^1$, I. Szleifer$^2$, M. Schick$^1$}
\affiliation{$^1$ Department of Physics, University of Washington,  Box
  351560, Seattle, WA 98195-1560 U. S. A.\\ 
       $^2$Department of Chemistry, Purdue University,
West Lafayette, IN 47907-1393 U. S. A.}

\date{\today, draft}

\begin{abstract} 
We employ a molecular model to study a ternary mixture 
of saturated lipid, with tails of sixteen carbons, a mono unsaturated 
lipid with tails of eighteen carbons, and cholesterol. The model, solved 
within mean-field theory, produces several forms of phase diagrams 
depending upon the relative strengths of interactions, but only one that 
shows the coexistence of two liquid phases observed in experiment. The 
lipids in the phase rich in cholesterol are more ordered than those in 
the other. The binary cholesterol, saturated lipid system also exhibits 
liquid, liquid coexistence. 
\end{abstract}
\maketitle

There has been enormous interest in the hypothesis that the lipids comprising
the plasma membrane are distributed inhomogeneously, with regions rich in
cholesterol and saturated lipids floating, like rafts, in a sea of unsaturated
lipids. Such rafts have been implicated in many biological processes including
endocytosis, transcription and transduction processes, and viral infection. The
extent of the interest is reflected in the number of recent
reviews \cite{simons00,edidin03,munro03,simons04,mcmullen04}. It has spurred
numerous {\em in vitro} investigation of ternary systems of cholesterol and a
high-melting point (hmp) lipid, either sphingomyelin or a saturated
glycerolipid, and a glycerolipid whose melting point is lower, either because
it is unsaturated \cite{dietrich01,veatch02,veatch03,almeida03} or because,
although saturated, it is relatively short \cite{silvius96,feigenson01}.  A
major question the ternary studies could address is whether the aggregation
phenomenon indicated {\em in vivo} might simply be attributed to liquid-liquid
phase separation. If so, it could be observed {\em in vitro}. The assumption
was that one fluid, the liquid disordered (ld) phase, would be rich in the
low-melting-point (lmp) lipid, while the other, the liquid ordered (lo) phase,
would be rich in cholesterol and the hmp lipid. The nomenclature ``liquid
ordered'' \cite{ipsen87} signifies that there are fewer thermally excited
gauche bonds in the hydrocarbon tails of the lipid chains in this phase than in
the liquid disordered phase. In addition to these two phases, a gel phase
occurs below the melting, or main chain transition temperature, of
the hmp lipid when the concentration of this lipid is large. This phase is the
most ordered of the three.

While there is some agreement on broad features in the interior of the ternary
diagrams, there is also considerable uncertainty. In particular the approach to
the cholesterol, hmp lipid binary system, and the phase diagram of that system
itself remain a matter of controversy. The disagreement centers on whether or
not there is coexistence of two liquid phases over some range of temperature
even in this binary system.

Evidence of such coexistence derives from several studies using different
probes: electron spin resonance \cite{shimshick73}, freeze fracture electron
microscopy \cite{lentz80}, nuclear magnetic resonance (NMR) \cite{vist90},
fluorescence resonance energy transfer \cite{loura01}, and other fluorescence
studies \cite{almeida03}. Perhaps the strongest evidence, however, is indirect,
coming from calorimetry \cite{vist90,mcmullen95}. In the cholesterol,
dipalmitoylphosphatidylcholine (DPPC) system, these measurements clearly show a
very large and narrow specific heat signal at a temperature which is rather constant over a
range of compositions, the standard indicator of a triple point. The signal was
so interpreted by Vist and Davis \cite{vist90}. Given the undisputed existence
of a gel phase which coexists with a liquid phase at low temperature, the
observation of a triple point leads to the conclusion of
the existence of a gel and two liquid phases at higher temperatures. There is
NMR evidence for such a triple point in another cholesterol, hmp lipid system
as well \cite{miao02}.
 
There are also arguments against the existence of two liquid phases in the cholesterol,
hmp lipid system. First, the sharp specific heat signals have been
interpreted as {\em not} arising from three-phase coexistence by McElhaney and
co-workers \cite{mcmullen95}, and they also dispute the conclusion drawn by
Vist and Davis from their NMR experiments.  Indeed the NMR work of Huang et al.
\cite{huang93} found no such coexistence, nor did those using fluorescence
recovery after photo bleaching \cite{rubenstein79}. Fluorescence
microscopy also does not observe phase separation in these systems
\cite{veatch02,veatch03}. 
Even if the behavior of this binary system were clear, its relation to that of
the canonical 1:1:1 raft composition would not be. The question would remain
whether the coexistence
of liquids in this part of the ternary system was directly related to
such coexistence in the binary system.

To clarify the ternary system and its evolution from the binary one, we have
carried out a study based upon a microscopic model. Previous theoretical
investigations of these mixtures have  been limited to the binary systems of
cholesterol and hmp lipid \cite{ipsen87,nielsen99,komura04}, or hmp and lmp
lipids \cite{elliott05}.  Notable among the former is the pioneering study of
Ipsen et al. which coined the description ``liquid-ordered phase''
\cite{ipsen87}.

Our ternary mixture consists of cholesterol, a lipid with two saturated tails
of sixteen carbons, C16:0, and another with two mono-unsaturated tails of
eighteen carbons, C18:1. There are essentially two adjustable interaction
parameters in the model, and their variation produces three distinct phase
diagrams. One is clearly irrelevant to experiment, one is consistent
with claims of no liquid, liquid coexistence in the binary system, and one
is consistent with claims that there is. Only the last of these exhibits
the liquid, liquid coexistence in the heart of the ternary diagram,
consistent with experiments.  

In our model, the lipids tails are described using the Flory rotational
isomeric states representation \cite{flory69}. One introduces the local volume
fraction per unit length for a lipid tail in a particular configuration, 
$\hat{\phi}_{\sigma}(z)$,
\begin{equation}
\hat{\phi}_{\sigma}(z)=\sum_{k=1}^{n_\sigma}\nu_{\sigma}(k)
\delta[z-z_{k}],
\label{phi}
\end{equation}
where the $z$ direction is perpendicular to the plane of the bilayer,
$\sigma={\rm s\ or\ u}$ is an index indicating whether the lipid is saturated
or unsaturated, the $\nu(k)=28$\AA$^3\equiv\nu_0$, the volume of the $k$'th
monomer if it is a CH$_2$, or 56\AA$^3$ if a CH$_3$ group, and $n_s=15$,
$n_u=17$. The local orientation of the chain is conveniently specified by the
normal to the plane determined by the $k$'th CH$_2$ group,
\begin{equation}
{\bf u}_{\sigma,k}=\frac{{\bf r}_{k-1}-{\bf r}_{k+1}}
{|{\bf r}_{k-1}-{\bf r}_{k+1}|}, \qquad k=1...n_s-1.
\label{normals}
\end{equation}
A similar local volume fraction is introduced for the cholesterol, with index
$\sigma=c$, $n_c=27$ and $\nu_c(k)\approx$21.0\AA$^3$. The orientation of the
small acyl chain of the cholesterol is specified in the same manner as are the
lipid chains, while the orientation of its rigid rings are specified by the
unit vector, ${\bf u_c}$, from the third to the seventeenth carbon in the
molecule, using the conventional labeling.

In earlier theories, \cite{gruen79,szleifer86}, the effect of the short-range
repulsive and long-range attractive interactions between elements was accounted
for approximately by replacing them by a constraint that the density within the
hydrophobic region be constant locally. The free energy of the system can then
be written in terms of the ${\hat\phi}$. While accounting well for many
properties of bilayers, this theory cannot lead to a main chain transition to a
gel phase. Hence we have added \cite{elliott05} an additional separable
interaction per unit volume which tends to align the elements with each other
and with the bilayer normal ${\bf c}$:
\begin{eqnarray}
V_{\sigma,\sigma'}&=&-(J_{\sigma,\sigma'}/\nu_0)g({\bf u}_{\sigma}\cdot{\bf c})
g({\bf u}_{\sigma'}\cdot{\bf c}),\\
g({\bf u}_{\sigma}\cdot{\bf c})&\equiv&(m+1/2)({\bf u}_{\sigma}\cdot{\bf
c})^{2m}
\end{eqnarray}
For large $m$, $g\approx m\exp(-m\theta^2)$ where $\theta$ is the angle between
the two unit vectors. Matching lipid parameters, we have taken
$m=18$. Note that this interaction between two vectors falls off exponentially
if either of them is not well aligned with the bilayer normal. We take the
strength of the local interactions between bonds in lipid tails to be the same,
$J_{ss}=J_{uu}=J_{su}\equiv J_{ll}$, irrespective of whether the bonds are in a
saturated or unsaturated chain, and the interaction strength between the
cholesterol and any bond in a lipid tail to be the same, $J_{sc}=J_{uc}\equiv
J_{lc}$, irrespective of the type of lipid chain. Thus the model contains three
interaction strengths: $J_{ll}$ which is set by the main chain transition
temperature, $J_{lc}$, and $J_{cc}$ the strength of the aligning interaction
between cholesterols.

Just as the ${\hat\phi}$ describe the local density, so we introduce a function
which describes the local ordering with respect to the bilayer normal
\cite{schmid98}:
\begin{equation} 
{\hat\xi}_{\sigma}(z)=\sum_{k=1}^{n_{\sigma}-1}\nu_{\sigma}(k) 
\delta(z-z_{k})g({\bf u}_{\sigma}\cdot{\bf c}). 
\end{equation} 
The Helmholtz free energy per unit area, $f_A(T,\rho_s,\rho_u,\rho_c)$, with
$\rho_{\sigma}=N_{\sigma}/A$, $\sigma=$ s or u, the areal density of saturated
or unsaturated chains, and $\rho_c=N_c/A$ the areal density of cholesterol, can
now be obtained directly within mean-field theory. One finds
\begin{eqnarray}
&\beta f_A&=-\frac{1}{\nu_0}\int\{(\beta/2)\sum_{\sigma,\sigma'}
J_{\sigma\sigma'}\rho_{\sigma}<{\hat\xi}_{\sigma}>\rho_{\sigma'}
           <{\hat\xi}_{\sigma'}>\nonumber\\
             &+&\sum_{\sigma}\rho_{\sigma}[<{\hat\xi}_{\sigma}>B_{\sigma}(z)+
                <{\hat\phi}_{\sigma}>\Pi(z)]\}dz\nonumber\\
                &-&\sum_{\sigma}\rho_{\sigma}\ln Q_{\sigma}+
                \frac{\rho_s}{2}\ln \frac{\rho_s}{\rho_0}+\frac{\rho_u}{2}\ln \frac{\rho_u}{\rho_0}
                +\rho_c\ln\frac{\rho_c}{\rho_0} \ \  
\end{eqnarray}
with $\rho_0\equiv V/A\nu_0$, $\beta\equiv 1/k_BT$.
Here $<O_{\sigma}(z)>=TrP_{\sigma}\hat O_{\sigma}(z)$, the average in a single
chain ensemble, with
\begin{eqnarray}
P_{\sigma}&=&\frac{1}{Q_{\sigma}}\exp\{-{\cal E}_{\sigma}\}
\equiv\frac{1}{Q_{\sigma}}\exp\{-\beta H_1\nonumber\\
&-&\frac{1}{\nu_0}\int[{\hat\phi}_{\sigma}(z)\Pi(z)+
{\hat \xi}_{\sigma}(z)B_{\sigma}(z)]dz\},\\
Q_{\sigma}&=&Tr\exp\{-{\cal E}_{\sigma}\},
\end{eqnarray}
and $H_1$ contains the intra chain energy arising from the presence of gauche
bonds. The three unknown fields $B_{s}=B_{u}\equiv B_{l}$, $B_{c}$, and $\Pi$
are obtained from minimization of the free energy with respect to the
$<{\hat\xi}_{\sigma}>$ and $<{\hat\phi}_{\sigma}>$ which leads to the three
local self consistent equations
\begin{eqnarray}
1&=&\sum_{\sigma}\rho_{\sigma}<{\hat\phi}_{\sigma}(z)>,\\ B_l&=&-\beta 
\{J_{ll}[\rho_s<{\hat\xi}_s>+\rho_u<{\hat\xi}_u>]+J_{lc}\rho_c<{\hat\xi}_c>\},
\nonumber\\ 
B_c&=&-\beta \{J_{lc}[\rho_s<{\hat\xi}_s>+\rho_u<{\hat\xi}_u>]+ 
J_{cc}\rho_c<{\hat\xi}_c>\}.\nonumber
\end{eqnarray}
The first of these is simply the incompressibility constraint. 

\begin{figure*}[t]
\includegraphics[width=6.5in,height=1.8in]{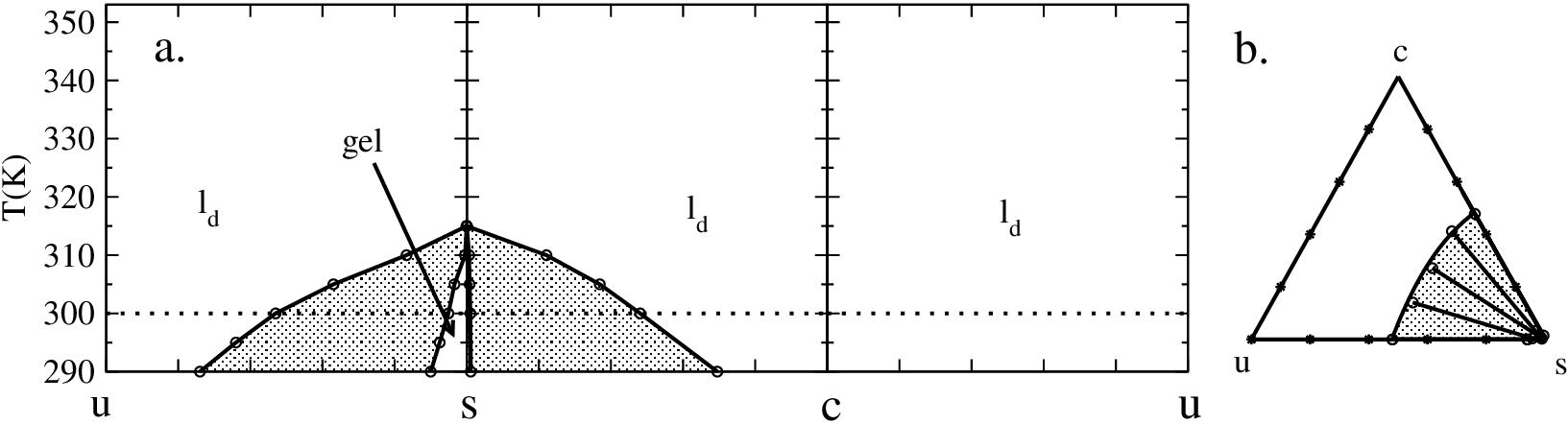}
\caption{Calculated phase diagrams of the three binary mixtures of cholesterol
(c), saturated (s), and unsaturated (u) lipids in temperature-composition space
for $J_{ll}(m+1/2)^2/k_BT^*=1.44$ and $J_{lc}=J_{cc}=0.0$.  These binary diagrams
form the sides of the Gibbs prism, a cut through which at 300K produces the
Gibbs triangle shown in Fig.  1b. Regions of two-phase coexistence are shaded,
and some tie lines are shown.}
\end{figure*}

The heart of the above method, and its difficulty, is the evaluation of the
single molecule partition functions $Q_{\sigma}$. To evaluate them, we have
generated on the order of $10^7$ configurations of each molecule.  Finally the
effect of the water, lipid interface is taken into account via a contribution
to the free energy per unit area of $\gamma_0$, set equal to the oil, water
tension. Thus the total free energy per unit area, or surface tension, is
$f_{tot}(T,\rho_s,\rho_u,\rho_c)=f_A+\gamma_0$. We look for phases for which
this surface tension vanishes. Phase equilibria is determined by standard
thermodynamic equalities \cite{elliott05}.
 
We now turn to the results of our calculation. The binary phase diagram of
unsaturated and saturated lipids \cite{elliott05} depends upon packing
constraints and the lipid-lipid interaction strength $J_{ll}$ which we set by
matching the main chain transition of the saturated lipid to $T^*=315K$,
that of DPPC. Below this temperature there is a gel phase, rich in the
saturated lipid, and a disordered liquid phase rich in the unsaturated lipid.
The main chain transition temperature of the unsaturated lipid is below
0$^{\circ}$C. The nature of the ternary diagram depends upon the relative
strengths, $J_{lc}$ and $J_{cc}$ of the lipid-cholesterol and
cholesterol-cholesterol aligning interactions. We find essentially three
classes of diagrams.

The first occurs when the other two interactions are weak compared to $J_{ll}$.
For the binary cholesterol, saturated lipid system, this results only in gel
and ld phases, and therefore no liquid, liquid coexistence, consistent with one
side of the controversy noted earlier.  The three binary phase diagrams as a
function of temperature and composition are shown in Fig. 1(a). A ternary
diagram at a temperature below the main chain temperature is shown in Fig.
1(b).  There is {\em no} region of liquid-liquid coexistence.  As such regions
can be seen directly in the experimental ternary systems
\cite{dietrich01,veatch02,veatch03}, this choice of interactions does not
apply.
\begin{figure*}[t]
\includegraphics[width=6.5in,height=1.8in]{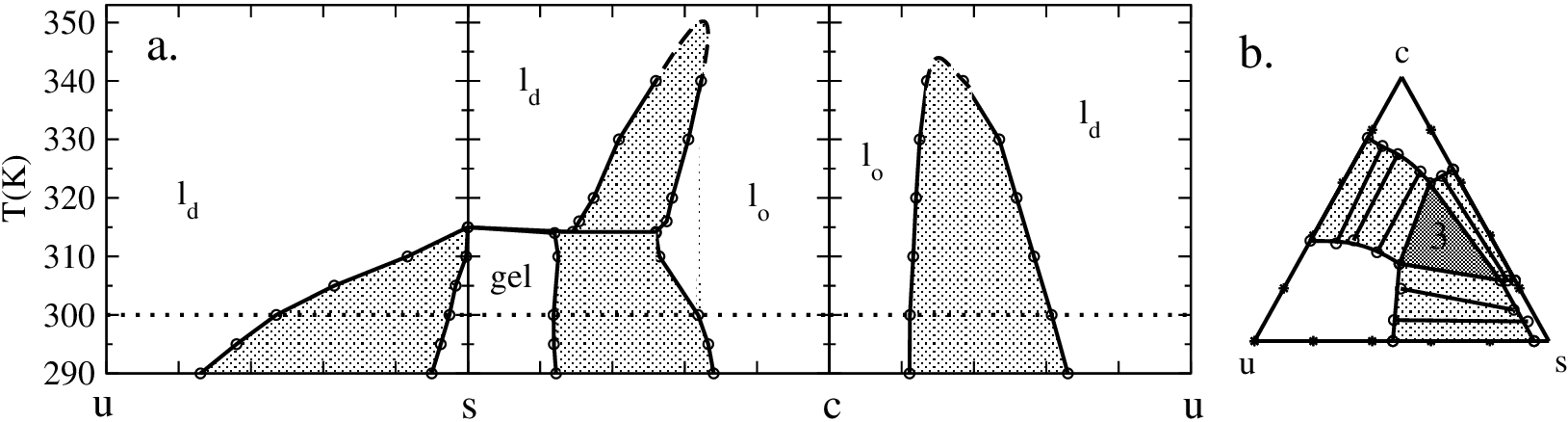}
\caption{ Figure 2a. Binary phase diagrams for $J_{lc}=0.78J_{ll}$ and
$J_{cc}=0.73J_{ll}$, and $J_{ll}$
as in Fig. 1. The saturated lipid-cholesterol mixture has a triple point
very near the main-chain transition temperature, so that the gel,
$l_d$ coexistence region is very narrow.  Dashed lines are extrapolations.
The ternary mixture at $T=300K$ is shown in Fig. 2b.\\}
\end{figure*}

A second class of diagram occurs if the lipid cholesterol aligning interaction,
$J_{lc}$, is dominant. This produces a gel phase which is swollen with
cholesterol. As there is no experimental evidence for this, we do not consider
this choice of interactions further.

A third class of diagram occurs when the interaction between cholesterols is
at least comparable to that between lipids. 
Binary phase diagrams based on our calculations are shown in Fig. 2(a).
In the cholesterol, saturated lipid system, there is coexistence of two liquid
phases over a range of high temperatures, and a temperature of three-phase
coexistence with a gel phase, consistent with refs
\cite{shimshick73}-\cite{loura01}. A cut through the ternary diagram at a
temperature of $T=300K$ below the triple temperature is shown in Fig. 2 (b).
There is a large region of coexistence of the lo, ld, and gel phases. There are
three regions of two-phase coexistence which extend to the binary axes. The
coexistence region between lo and ld liquids would be identified as the region
in which rafts could exist.  The diagram is topologically identical to that
reported in ref \cite{almeida03}. 

The degree of order, $S_{\rm CD}$, in the lipid chains is given in terms of the second
Legendre polynomial $|S_{\rm CD}|=|P_2(\theta_k)|/2$, 
where $\theta_k$ is the angle between
the normals to the bilayer and to the plane of the $k$'th CH$_2$ group. The
order parameter for the saturated lipids is shown in Fig. 3 for each of the
three phases which coexist in the phase diagram of Fig. 2b. As can be seen, the
chains in the lo phase are indeed more ordered than those in the ld phase.
This increased order is due to the interactions of the lipids with cholesterol,
which is itself well ordered. We have also verified that the addition of
cholesterol tends to disorder the gel phase, in agreement with experiment
\cite{vist90}.
\begin{figure}
\begin{center}
\includegraphics[scale=0.35]{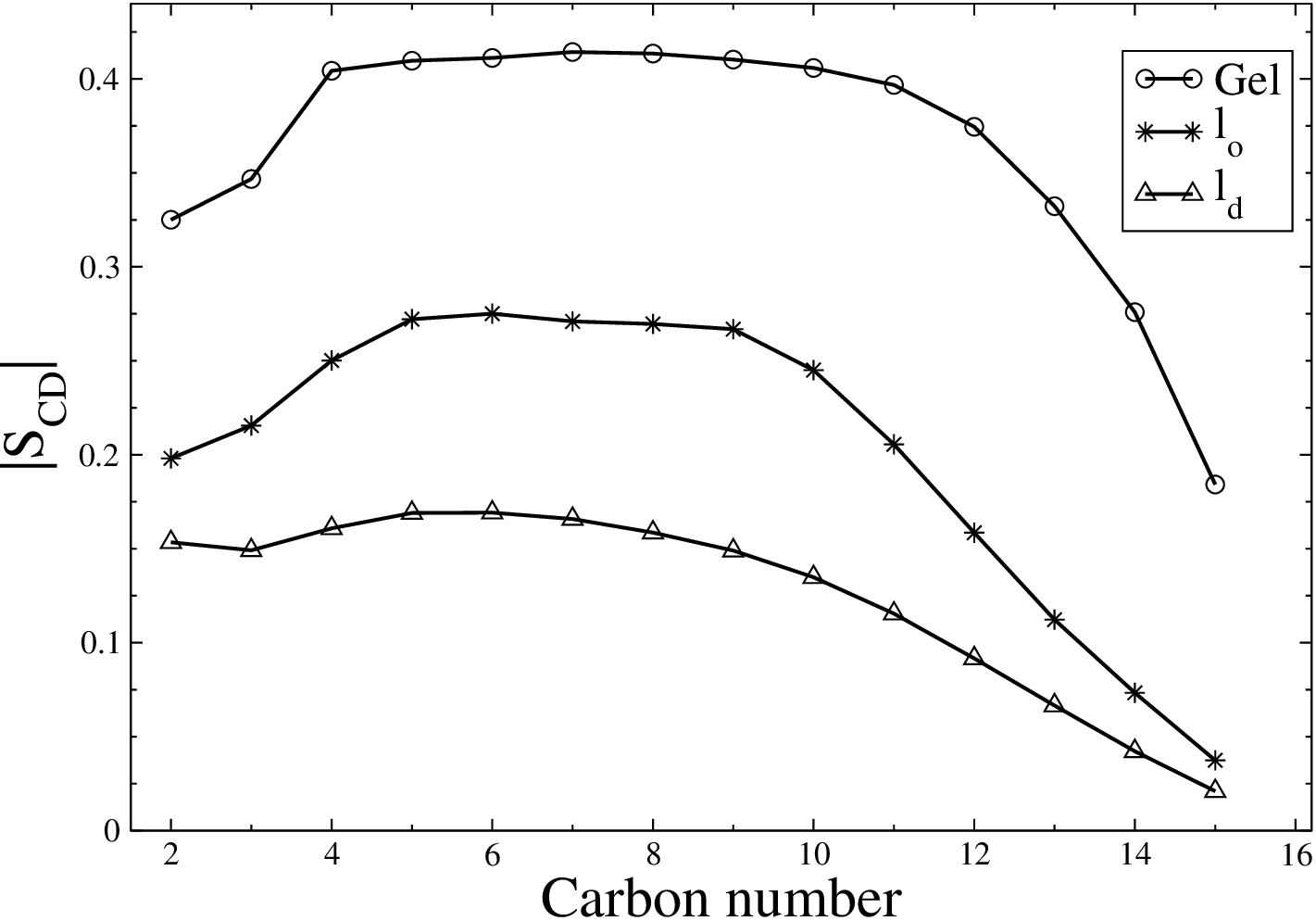}
\caption{The order parameters for the saturated tails in the three coexisting
phases of the ternary mixture at 300K.  Cholesterol concentration is
greatest in the lo
phase, which is more ordered than the ld phase. The effect of the larger
cholesterol concentration is evident in the plateau corresponding to ordering
about cholesterol's ring structure.} 
\end{center}
\end{figure}

The evolution of the ternary diagram with increasing temperature is simple.
The region of three-phase coexistence shrinks and vanishes at the triple
temperature of the binary cholesterol, hmp lipid system. Above the main chain
temperature of this lipid, there is a two-phase region of lo and ld coexistence
which stretches from one binary system with cholesterol to the other. At higher
temperatures, this region detaches from the binary cholesterol, lmp lipid axis
at a critical point, and the region shrinks with further increase of
temperature until it vanishes at a critical point on the cholesterol, hmp lipid
axis. It is clear from the evolution of our calculated ternary diagram that the
liquid, liquid coexistence within our ternary diagram is intimately tied to a
similar coexistence in the binary cholesterol, saturated lipid system. Given
that the coexistence of liquids has been observed in the ternary systems, the
calculation gives strong support to the view that such coexistence already
exists in the cholesterol, hmp lipid system as well. 

We are grateful to Sarah Keller and Sarah Veatch for numerous profitable
discussions. This work was supported by the National Science Foundation under
Grants No. DMR-0140500 and CTS-0338377.

\bibliography{lipids4}

\end{document}